\documentclass{ws-ijqi}
\renewcommand{\draftnote}{\centerline%
{\underline{\texttt{"\jobname.tex", draft of \today, \currenttime}}}}

\renewcommand{\draftnote}{\relax}

\def\currenttime{\hour=\time \divide\hour by 60 \number\hour:%
  \multiply\hour by 60 \minute=\time \global\advance\minute by -\hour%
  \ifnum\minute<10 0\number\minute\else\number\minute\fi}

\newtheorem{Thm}{Theorem}

\newtheorem{Rmk}{Remark}
\newtheorem{Epl}{Example}
\newcommand{\fnmark}[1]{$^{\ref{#1}}$}

\newcommand{\eprint}[2][quant-ph]{\mbox{e-print} #1/\linebreak[0]#2}

\begin{document}

\markboth{B.-G. Englert, F.-W. Fu, H. Niederreiter, and C. Xing}
{Codes for Key Generation in Quantum Cryptography}


\title{
\uppercase{Codes for Key Generation\\ in Quantum Cryptography}}

\author{\uppercase{Berthold-Georg Englert}}
\address{Department of Physics, %
National University of Singapore, Singapore 117542\\
phyebg@nus.edu.sg}

\author{\uppercase{Fang-Wei Fu}%
\thanks{On leave from the Department of Mathematics, %
Nankai University, Tianjin 300071, P. R. China}}
\address{Temasek Laboratories, %
National University of Singapore, Singapore 117508\\
tslfufw@nus.edu.sg}

\author{\uppercase{Harald Niederreiter} and \uppercase{Chaoping Xing}}
\address{Department of Mathematics, %
National University of Singapore, Singapore 117543\\
nied@math.nus.edu.sg, matxcp@nus.edu.sg}

\maketitle

\begin{history}
\received{12 April 2005}
\end{history}

\begin{abstract}
As an alternative to the usual key generation by two-way communication in
schemes for quantum cryptography, we consider codes for key generation by
one-way communication.
We study codes that could be applied to the raw key sequences that are
ideally obtained in recently proposed scenarios for quantum key distribution, 
which can be regarded as communication through symmetric four-letter channels.
\end{abstract}

\keywords{Error correcting codes, linear codes, quantum key distribution}

\section{Introduction} \label{sec1}

In a recently proposed protocol for quantum key
distribution,\cite{Renes,TetraCrypt} Alice sends uncorrelated qubits through a
quantum channel to Bob.
Under ideal circumstances, the channel is noiseless, and then the situation is
as follows.

Alice prepares each qubit in one of four states ---
labeled $A$, $B$, $C$, and $D$, respectively, and chosen at random ---
and Bob detects each qubit in one of four states that are labeled
correspondingly.
The set-up has the peculiar feature that Bob \emph{never} obtains the letter
that specifies the state prepared by Alice.
Rather, he always gets one of the other three letters, whereby the laws of
quantum physics ensure that the outcome is truly random, and each possibility
occurs equally likely.

These physical laws also prevent any third party, eavesdropper Eve,
from acquiring information about Alice's or Bob's letters.
Therefore, they can exploit the correlations between their letters to generate
a private cryptographic key, which they can then use for the secure encryption
of a message.

The key generation is a crucial step.
Two different procedures are described in Refs.~\refcite{Renes} and
\refcite{TetraCrypt}, with respective efficiencies of $\frac{1}{3}$ and
$\frac{2}{5}$ key bits per letter.
Both procedures rely on \emph{two-way} communication between Alice and Bob.
By contrast, it is our objective here to study codes for the key generation by
\emph{one-way} communication.

After the exchange of many qubits through the quantum channel, Alice and Bob
have random sequences of the four letters, such that corresponding letters
are never the same, while each of the twelve pairs of different letters occurs
one-twelfth of the time, with no correlations between the pairs.
Alice sends a code word to Bob by telling him, through a public channel, the
positions at which the letters appear in her sequence --- such as
``3rd letter, then 14th, 15th, 92nd, and 65th'' for a particular five-letter
word.
Bob forms the received word from his corresponding letters, and then decodes.

The public communication does not leak any useful information to Eve.
Thus, if Alice chooses a random sequence of code words,
each word being equally likely, as she will do,
Eve knows nothing about Alice's words.
She also knows nothing about Bob's decoded words, provided
that Bob's decoding procedure does not favor some words at the expense
of others.
Accordingly, the sequence of words constitutes a privately shared key for
secure classical communication between Alice and Bob.

There is a nonzero probability that Bob's received word is consistent with two
or more words that Alice could have sent, so that the decoding will not be
completely error-free.
A good, practical code must, therefore, represent a compromise between (i)
having not too many code words, (ii) an acceptable error rate, and (iii) a
reasonable efficiency.
Arguably the best compromise we report in Section~\ref{sec8} is code~(3) of
Example~1.
It has 1024 words, an error rate of 0.6\%, and an efficiency of $\frac{1}{4}$ 
key bits per letter.

As there is no fundamental reason why the key generation by one-way
communication should be substantially less efficient than that by two-way
communication, one expects that more efficient codes can be found.
Therefore, the work reported here should be regarded as a first step, not as
the final word on the matter. 

It is worth mentioning that there is a very similar problem for the
three-letter channel of Renes's ``trine'' scheme.\cite{Renes}
Further, the standard BB84 protocol\cite{BB84} has a four-letter channel with
quite different properties, for which codes for one-way key generation are not
known.
The same remark applies to the six-letter generalization\cite{6states} of BB84.
In short, there is a whole class of coding problems that deserve attention.

\section{Probability Distributions} \label{sec2}

The quantum protocols of Alice and Bob involve two
random variables $X$ and $Y$ taking values in $\{A, B, C, D\}$.
We have the following corresponding probability distributions with
$x,y \in \{A,B,C,D\}$. The joint probability distribution of $X$ and $Y$ is
given by
\begin{equation}
\Pr\{X=x, Y=y\} = \left\{
\begin{array}{cl}
0 & \mbox{if}\ y=x\,, \\[1ex]
\displaystyle\frac{1}{12} & \mbox{if}\ y\not= x\,.
\end{array}
\right. \label{a1}
\end{equation}
Accordingly, the marginal probability distributions of $X$ and $Y$ are 
\begin{equation}
\Pr\{X=x\} = \frac{1}{4}\,, \qquad \Pr\{Y=y\} = \frac{1}{4}\,, \label{a2}
\end{equation}
and the conditional probability distribution of $Y$ with respect to $X$ is
\begin{equation}
\Pr\{Y=y\;|\;X=x\} = \left\{
\begin{array}{cl}
0    & \mbox{if}\ y=x\,, \\[1ex]\displaystyle
\frac{1}{3} & \mbox{if}\ y\not= x\,.
\end{array}
\right. \label{a3}
\end{equation}

Now we compute the information-theoretic quantities entropy, conditional
entropy, and mutual information%
\footnote{\label{fn1}For the definitions of these and other
information-theoretic quantities see Ref.~\refcite{cover}, for example.}  
of the random variables $X$ and $Y$.
The entropy of $Y$ is 
\begin{equation}
H(Y)= \mathop{- \sum}_{y\in \{A,B,C,D\}}\hspace{-1em}
\Pr\{Y=y\} \log_{2} \Pr\{Y=y\} =
2\,, \label{a4}
\end{equation}
and for the conditional entropy of $Y$ with respect to $X$ we find
\begin{equation}
H(Y | X)= 
\mathop{-\sum}_{\mbox{\scriptsize$\begin{array}{c} x, y\in \{A,B,C,D\}\\%
 y \neq x\end{array}$}}\hspace{-2em}
\Pr\{X=x, Y=y\} \log_{2}\Pr\{Y=y | X=x\} = \log_{2}3\,, 
\label{a5}
\end{equation}
and we obtain 
\begin{eqnarray}
I(X; Y) & = & 
\hspace{-1em}
\sum_{\mbox{\scriptsize$\begin{array}{c} x, y\in \{A,B,C,D\}\\%
 y \neq x\end{array}$}}\hspace{-2em}
\Pr\{X=x, Y=y\} \log_{2}
\frac{\Pr\{X=x, Y=y\}}{\Pr\{X=x\} \Pr\{Y=y\}}  \nonumber\\
& = & H(Y) - H(Y|X) = \log_{2} \frac{4}{3} \label{a6}
\end{eqnarray}
for the mutual information of $X$ and $Y$.

\section{Discrete Memoryless Channel} \label{sec3}

The information transmission from Alice to Bob can be described in
information theory by a discrete memoryless channel.\fnmark{fn1}
This channel is
characterized by the conditional probability distribution\fnmark{fn1} of $Y$
with respect to $X$,
\begin{equation}
Q(y|x)=\Pr\{Y=y\;|\;X=x\} = \left\{
\begin{array}{cl}
0    & \mbox{if}\ y=x\,, \\[1ex]\displaystyle
\frac{1}{3} & \mbox{if}\ y\not= x\,,
\end{array}
\right. \label{a7}
\end{equation}
where $x,\; y \in \{A, B, C, D\}$.
The \emph{channel capacity}\fnmark{fn1} is defined by
\begin{equation}
c= \max_{P_{X}} I(X; Y) = \max_{P} I(P;Q)\,, \label{a8}
\end{equation}
where
\begin{equation}
I(P; Q)  =  \sum_{x, y} P(x)Q(y|x) \log_{2}
\frac{Q(y|x)}{\sum_{x'}P(x')Q(y|x')} \label{a9}
\end{equation}
and the maximum is taken over all probability distributions $P$ on
$\{A, B, C, D\}$.

The channel defined by (\ref{a7}) is a symmetric channel (see
Theorem 8.2.1 on p.~190 in Ref.~\refcite{cover}). 
Hence, the capacity is attained by the uniform distribution 
on $\{A, B, C, D\}$, so that
\begin{equation}
c = \log_{2} \frac{4}{3}\doteq0.4150\,. \label{a10}
\end{equation}
For any positive integer $n$, the $n$th extension of this discrete
memoryless channel has the conditional probability distribution
\begin{equation}
Q^{n}(\mathbf{y} | \mathbf{x})= \left\{
\begin{array}{cl}
0    & \mbox{if $y_i =x_i$ for some $i$,} \\[1ex]\displaystyle
\frac{1}{3^{n}} & \mbox{if $y_i \neq x_i$  for all $i$},
\end{array}
\right. \label{a11}
\end{equation}
where $\mathbf{x}=(x_1,x_2,\ldots,x_n)$, $\mathbf{y}=(y_1,y_2,\ldots,y_n) \in
\{A,B,C,D\}^n$ are $n$-letter words.

\section{Codes for the Specific Channel} \label{sec4}

In this section, we discuss the design of codes and decoding methods for
the specific channel introduced in Section \ref{sec3}.

Let $\mathbf{F}_{4}$ be the finite field with four elements. It is
convenient to let $A$, $B$, $C$, $D$ be represented respectively 
by the four elements $0$, $1$, $a$, $b$ of $\mathbf{F}_{4}$
since we want to use linear codes for this specific channel.
The addition and multiplication tables of $\mathbf{F}_4$ are as follows:
\begin{equation}
  \begin{array}[c]{c|cccc}
+ & 0 & 1 & a & b \\ \hline
0 & 0 & 1 & a & b \\
1 & 1 & 0 & b & a \\
a & a & b & 0 & 1 \\
b & b & a & 1 & 0
  \end{array}
\qquad\qquad
  \begin{array}[c]{c|cccc}
\times & 0 & 1 & a & b \\ \hline
     0 & 0 & 0 & 0 & 0 \\
     1 & 0 & 1 & a & b \\
     a & 0 & a & b & 1 \\
     b & 0 & b & 1 & a
  \end{array}
\end{equation}

Let $\mathbf{F}_{4}^{n}$ be the $n$-dimensional vector space 
over $\mathbf{F}_{4}$. For two vectors
$\mathbf{x}=(x_1, x_2, \ldots, x_n) \in \mathbf{F}_{4}^{n}$, 
$\mathbf{y}=(y_1, y_2, \ldots, y_n) \in \mathbf{F}_{4}^{n}$, the \emph{Hamming
distance} $d(\mathbf{x}, \mathbf{y})$ between $\mathbf{x}$ and $\mathbf{y}$ is
defined as the number of coordinates in which they differ, 
\begin{equation}
d(\mathbf{x}, \mathbf{y}) = \bigl| \{i: x_i \not= y_i\} \bigr|\,.  
\end{equation}
The \emph{Hamming weight} $w(\mathbf{x})$ is  the number of nonzero
coordinates in $\mathbf{x}$, 
\begin{equation}
w(\mathbf{x}) = \bigl| \{i: x_i \not= 0\} \bigr|\,.  
\end{equation}

A code of length $n$ with $M \ge 2$ codewords is a subset $\mathcal{C}$ of
$\mathbf{F}_{4}^{n}$,
\begin{equation}
\mathcal{C} = \{ \mathbf{c}_1, \mathbf{c}_2, \ldots, \mathbf{c}_M \}\,, 
\qquad
\mathbf{c}_i \in \mathbf{F}_{4}^{n}\,. \label{b1}
\end{equation}
The \emph{minimum distance} $d(\mathcal{C})$ of the code $\mathcal{C}$ is the
minimum Hamming distance between two distinct codewords, 
\begin{equation}
d(\mathcal{C}) 
= \min\{ d(\mathbf{x}, \mathbf{y}): \mathbf{x}, \mathbf{y} \in \mathcal{C}, 
\; \mathbf{x}\not=\mathbf{y} \}\,. \label{bb1}
\end{equation}

We denote by
\begin{equation}
L(\mathbf{c}_i) = \{ \mathbf{x} \in \mathbf{F}_{4}^{n}: \; d(\mathbf{x}, 
\mathbf{c}_i)=n \}\quad\mbox{for $i=1,2, \ldots, M$}
\label{b2}
\end{equation}
the set of $n$-letter words that could be received if codeword $\mathbf{c}_i$
is sent. 
It is easy to see that for any $i, j, l \in \{1,2,\ldots,M\}$ we have
\begin{eqnarray}
& & \bigl|L(\mathbf{c}_i)\bigr| = 3^{n}\,,
\label{l1}\\
& & \bigl|L(\mathbf{c}_i) \cap L(\mathbf{c}_j)\bigr| \geq 2^{n}\,,
\label{l2}\\
& & \bigl|L(\mathbf{c}_i) \cap L(\mathbf{c}_j) \cap L(\mathbf{c}_l)\bigr| 
\geq 1\,.
\label{l3}
\end{eqnarray}
Note, in particular, the significance of (\ref{l3}): 
For any three different codewords that could have been sent by Alice, there is
at least one word received by Bob that is consistent with all three.  

Further, it follows from (\ref{a11}) that
\begin{equation}
Q^{n}(\mathbf{y} | \mathbf{x})= \left\{
\begin{array}{cl}
0    & \mbox{if}\ \mathbf{y}\not\in L(\mathbf{x})\,, \\[1ex]\displaystyle
\frac{1}{3^{n}} & \mbox{if}\ \mathbf{y}\in L(\mathbf{x})\,.
\end{array}
\right. \label{bb2}
\end{equation}
This means that if Alice sends $\mathbf{x} \in \mathbf{F}_{4}^{n}$ through
this channel to Bob, then Bob receives $\mathbf{y}\in L(\mathbf{x})$ with
probability $1/3^n$.

Now we describe a decoding method for the code $\mathcal{C}$ by
decoding regions, which exploits the significance of $L(\mathbf{c}_i)$.
The $M$ subsets $D_1, D_2, \ldots, D_M$ of $\mathbf{F}_4^n$ are called
\emph{decoding regions} for the code $\mathcal{C}$ if they satisfy 
the following conditions:
\begin{equation}
  \begin{array}{r@{\ }l}
\textrm{(i)} &D_i \subseteq L(\mathbf{c}_i)\,,\quad i=1,2, \ldots, M\,;
\\
\textrm{(ii)} &D_i \cap D_j = \emptyset\,,\quad i\not=j\,;
\\
\textrm{(iii)} &
\displaystyle\bigcup_{i=1}^{M} D_i =  \bigcup_{i=1}^{M} L(\mathbf{c}_i)\,.
  \end{array}
\end{equation}
The decoding method for the code $\mathcal{C}$ with the decoding
regions $D_1, D_2, \ldots, D_M$ is then: decode the received
vector $\mathbf{y}$ into $\mathbf{c}_i$ if $\mathbf{y} \in D_i$.
In some cases, for simplicity, we can construct the decoding regions in
accordance with
\begin{eqnarray}
D_1 & = & L(\mathbf{c}_1)\,,\nonumber\\ 
D_2 &=& L(\mathbf{c}_2) \setminus L(\mathbf{c}_1)\,,
\nonumber\\
D_i & = & L(\mathbf{c}_i)\setminus\left(\bigcup_{j=1}^{i-1} 
L(\mathbf{c}_j)\right)\,,\quad i=2,3, \ldots, M\,. \label{b3}
\end{eqnarray}
This means that we decode the received vector 
$\mathbf{y} \in \bigcup_{i=1}^{M} L(\mathbf{c}_i)$ into the first 
$\mathbf{c}_i$ such that $d(\mathbf{y}, \mathbf{c}_i)=n$.

Since the decoding regions of (\ref{b3}) refer to an agreed-upon order of the
codewords, the decoding is biased toward the early codewords in the list at 
the expense of the later ones.
Such a bias is avoided by the \emph{maximum likelihood decoding}.\cite{ga}
It can be described as follows: The received vector $\mathbf{y}
\in \bigcup_{i=1}^{M} L(\mathbf{c}_i)$ is decoded into any codeword
$\mathbf{c}_i$ such that $Q^{n}(\mathbf{y} | \mathbf{c}_i)$ is the maximum
value of $Q^{n}(\mathbf{y} | \mathbf{c})$ over all codewords 
$\mathbf{c}\in\mathcal{C}$. 
If there is more than one such $\mathbf{c}_i$, 
we choose one of them at random.
It is easy to see that this is equivalent
to the following decoding method: The received vector $\mathbf{y}$ 
is decoded into either one of the codewords $\mathbf{c}_i$ that
obey $d(\mathbf{y}, \mathbf{c}_i)=n$, choosing one at random if there are
several such $\mathbf{c}_i$s.

\section{Decoding Error Probability} \label{sec5}

In this section, we discuss the decoding error probability and Shannon's
Channel Coding Theorem for the specific channel introduced in Section
\ref{sec3}. Some criteria for good codes for this channel are given.

For a code $\mathcal{C}$ with decoding regions $D_1, D_2, \ldots, D_M$,
the probability $e_i$ of the event that the vector $\mathbf{y}$ received by
Bob is not decoded into the codeword $\mathbf{c}_{i}$ sent by Alice
is given by
\begin{eqnarray}
e_{i}  =  {\rm Pr}\{ \mathbf{y} \not\in D_{i} \mid \mathbf{c}_{i} \;
\mbox{is sent} \} 
& = & 1 - {\rm Pr}\{ \mathbf{y} \in D_{i} \mid \mathbf{c}_{i} \;
\mbox{is sent} \} \nonumber\\
& = & 1 - \frac{|D_i|}{3^n}\,,\quad i=1,2, \ldots, M\,. \label{b4}
\end{eqnarray}
The \emph{average error probability} $\bar{e}$ is the arithmetic mean of the
$e_i$s, 
\begin{eqnarray}
\bar{e} = \frac{1}{M} \sum_{i=1}^{M} e_{i}  
& = & 1 - \frac{1}{3^{n}M} \sum_{i=1}^{M} |D_i|  \nonumber\\
& = & 1 - \frac{1}{3^{n}M} \left|\bigcup_{i=1}^{M} D_i\right|  \nonumber\\
& = & 1 - \frac{1}{3^{n}M} \left|\bigcup_{i=1}^{M} L(\mathbf{c}_i)\right|\,,
\label{b5}
\end{eqnarray}
and the \emph{maximum error probability} $e_{\max}$ is the largest one 
of them,
\begin{equation}
e_{\max}= \max_{1\leq i \leq M} e_{i}. \label{b6}
\end{equation}
Obviously, $\bar{e}\leq e_{\max}$. Note that
\begin{eqnarray}
\bar{e}=0 & \Longleftrightarrow & e_{\max}=0 \nonumber\\
& \Longleftrightarrow & e_i =0 \ \mbox{for all $i$} \nonumber\\
& \Longleftrightarrow & L(\mathbf{c}_i) \cap L(\mathbf{c}_j) =
\emptyset\,,\quad i\neq j\,.
\end{eqnarray}
Hence, it follows from (\ref{l2}) that
\begin{equation}
e_{\max} \geq \bar{e} >0\,.  
\end{equation}
In particular, if the decoding regions $D_1, D_2, \ldots, D_M$
are given by (\ref{b3}), then
\begin{eqnarray}
e_{1} & = & 1 - \frac{1}{3^n}\bigl|L(\mathbf{c}_1)\bigr|=0\,, \nonumber\\
e_{i} & = & 1 - \frac{1}{3^n}\left|L(\mathbf{c}_i) \setminus
\textstyle\left(\bigcup_{j=1}^{i-1} L(\mathbf{c}_j) \right)\right|\nonumber\\
& = & \frac{1}{3^n}\biggl[\bigl|L(\mathbf{c}_i)\bigr|
-\left|\textstyle L(\mathbf{c}_i) \setminus \left(\bigcup_{j=1}^{i-1}
L(\mathbf{c}_j) \right)\right|\biggr]
\nonumber\\
& = & \frac{1}{3^n}
\left|\textstyle L(\mathbf{c}_i) \cap \left(\bigcup_{j=1}^{i-1} 
L(\mathbf{c}_j)\right)\right|
\nonumber\\
& = & 
\frac{1}{3^n}\textstyle\left|\bigcup_{j=1}^{i-1} \Bigl(L(\mathbf{c}_i)
\cap L(\mathbf{c}_j)\Bigr)\right|\,,\quad i=2,3,\ldots,M\,.
\label{b7}
\end{eqnarray}

The decoding error probability of a code is one of its most
important performance characteristics. In this connection, we recall
Shannon's Channel Coding
Theorem (see Refs.~\refcite{cover} and \refcite{ga}).

\begin{Thm}
\textbf{(Shannon's Channel Coding Theorem)}
For any $0<\varepsilon<1$ and $0<R<\log_{2}(4/3)$, there exists for
sufficiently large $n$ a code $\mathcal{C}$ of length $n$ and size
$M\doteq 2^{nR}$ such that $e_{\max} \leq \varepsilon$, and so in
particular $\bar{e}\leq \varepsilon$.
\label{theo-shannon}
\end{Thm}

\begin{Rmk}
For fixed $n$ and $M$, the best we can do is to choose a code
\begin{equation}
\mathcal{C} = \{ \mathbf{c}_1, \mathbf{c}_2, \ldots, \mathbf{c}_M \}\,, 
\qquad
\mathbf{c}_i \in \mathbf{F}_{4}^{n}\,, 
\end{equation}
such that $\bigl|\bigcup_{i=1}^{M} L(\mathbf{c}_i)\bigr|$ 
is as large as possible, i.e., the average error probability
$\bar{e}$ is as small as possible.
For fixed $n$ and $\bar{e}\leq \varepsilon$, in view of (\ref{b5}) we
have to try to find a code
$\mathcal{C}$ with the largest size $M$ such that
\begin{equation}
\frac{1}{M}\left|\bigcup_{i=1}^{M} L(\mathbf{c}_i)\right|
\geq 3^{n}(1-\varepsilon)\,.  
\end{equation}
Note that for certain values of $n$ and $\varepsilon$, 
such a code $\mathcal{C}$ may not exist.
\end{Rmk}

\section{Upper Bounds on the Decoding Error Probability} \label{sec6}

In this section, we give several upper bounds on the decoding error
probability of codes for our specific channel.

Let $\mathcal{C}$ be a code for our specific channel. 
The \emph{distance distribution} of the code $\mathcal{C}$ is defined by
\begin{equation}
A_{s} = \frac{1}{M} \bigl|\{ (\mathbf{x}, \mathbf{y})\in 
\mathcal{C} \times \mathcal{C}: d(\mathbf{x}, \mathbf{y})=s \}\bigr|\,, 
\quad s=0,1, \ldots, n\,. \label{c1}
\end{equation}
It is easy to see that
\begin{equation}
A_{0}=1\,,\quad \sum_{s=0}^{n} A_s = |\mathcal{C}|=M\,.
\label{cc1}
\end{equation}

\begin{Thm}
Let $\mathcal{C}$ be a code for the specific channel in Section \ref{sec3}.
Suppose that the distance
distribution of the code $\mathcal{C}$ is given by $A_0, A_1, \ldots,
A_n$. Then the average error probability $\bar{e}$ is upper bounded by
\begin{equation}
\bar{e} \leq \frac{1}{2} \sum_{s=1}^n A_s
\left(\frac{2}{3}\right)^s\,. \label{c2}
\end{equation}
\label{theo1}
\end{Thm}

\begin{proof} By the definition of $L(\mathbf{c}_i)$ in (\ref{b2})
we know that if $d(\mathbf{c}_i, \mathbf{c}_j)=s$, then
\begin{equation}
\bigl|L(\mathbf{c}_i) \cap L(\mathbf{c}_j)\bigr| 
= 3^{n-s}2^{s}= 3^{n} \left(\frac{2}{3}\right)^{s}\,,
\quad s=1,2, \ldots, n\,. \label{c3}
\end{equation}
Hence,
\begin{equation}
\sum_{1\leq i<j\leq M} \bigl|L(\mathbf{c}_i) \cap L(\mathbf{c}_j)\bigr| =
\frac{3^{n}M}{2} \sum_{s=1}^{n} A_s \left(\frac{2}{3} \right)^{s}\,. 
\label{c4}
\end{equation}
By (\ref{c4}) and noting that $|L(\mathbf{c}_i)|=3^{n}$, we obtain
\begin{eqnarray}
\left|\bigcup_{i=1}^{M} L(\mathbf{c}_i)\right| 
& \geq & \sum_{i=1}^{M} |L(\textbf{c}_i)| -
\sum_{1\leq i<j\leq M} \bigl|L(\mathbf{c}_i) \cap L(\mathbf{c}_j)\bigr| 
\nonumber\\
& =& 3^{n}M-\frac{3^{n}M}{2}\sum_{s=1}^{n}A_s\left(\frac{2}{3}\right)^{s}\,.
\label{c5}
\end{eqnarray}
Hence, (\ref{c2}) follows from (\ref{b5}) and (\ref{c5}). 
\end{proof}

\begin{Thm}
Let $\mathcal{C}$ be a code for the specific channel in Section \ref{sec3}.
If the minimum distance $d(\mathcal{C})\geq d$, 
then the average error probability $\bar{e}$ is upper bounded by
\begin{equation}
\bar{e} \leq \frac{M-1}{2} \left(\frac{2}{3}\right)^d <
\frac{M}{2} \left(\frac{2}{3}\right)^d\,. \label{c6}
\end{equation}
\label{theo2} 
\end{Thm}

\begin{proof} If the minimum distance $d(\mathcal{C})\geq d$, then
$A_s=0$ for $1\leq s \leq d-1$. Hence, by Theorem \ref{theo1} and
(\ref{cc1}),
\begin{equation}
\bar{e} \leq \frac{1}{2} \left(\frac{2}{3}\right)^d \sum_{s=d}^n
A_s = \frac{M-1}{2} \left(\frac{2}{3}\right)^d < \frac{M}{2}
\left(\frac{2}{3}\right)^d\,.
\end{equation}
This completes the proof. 
\end{proof}

\begin{Thm}
Let $\mathcal{C}$ be a code for the specific channel in Section \ref{sec3}.
If the minimum
distance $d(\mathcal{C})\geq d$ and the decoding regions $D_1, D_2,
\ldots, D_M$ are given by (\ref{b3}), then the maximum error
probability $e_{\max}$ is upper bounded by
\begin{equation}
e_{\max} \leq (M-1) \left(\frac{2}{3} \right)^d < M \left(\frac{2}{3}
\right)^d\,. \label{c7}
\end{equation}
\label{theo3}
\end{Thm}

\begin{proof} If the minimum distance $d(\mathcal{C})\geq d$, then for
$i\not=j$,
\begin{equation}
\bigl|L(\mathbf{c}_i) \cap L(\mathbf{c}_j)\bigr|  = 
3^{n-d(\mathbf{c}_i, \mathbf{c}_j)}2^{d(\mathbf{c}_i, \mathbf{c}_j)}
 =  3^{n} \left(\frac{2}{3} \right)^{d(\mathbf{c}_i, \mathbf{c}_j)}
\leq  3^{n} \left(\frac{2}{3} \right)^{d}\,. \label{c8}
\end{equation}
It follows from (\ref{c8}) that for $i=2, 3, \ldots, M$,
\begin{eqnarray}
\left|\bigcup_{j=1}^{i-1} \bigl(L(\mathbf{c}_i) \cap L(\textbf{c}_j)\bigr)
\right| \leq \sum_{j=1}^{i-1} \bigl|L(\mathbf{c}_i) \cap
L(\mathbf{c}_j)\bigr| &\leq& (i-1) 3^{n} \left(\frac{2}{3} \right)^{d} 
\nonumber\\ &\leq&  (M-1) 3^n \left(\frac{2}{3} \right)^d\,.
\label{c9}
\end{eqnarray}
Hence, by (\ref{b7}) and (\ref{c9}),
\begin{equation}
e_{1}=0\,,\quad e_i \leq (M-1) \left(\frac{2}{3} \right)^d\,,\quad
i=2,3,\ldots, M\,.  
\end{equation}
Therefore, $e_{\max} \leq (M-1) (2/3)^d$. 
\end{proof}

\section{Linear Codes} \label{sec7}

In this section, we discuss the design of linear codes and the decoding
error probability of linear codes for our specific channel. Some
criteria for good linear codes are given.

First, we recall some basic concepts for linear codes. A code ${\cal
C}$ over $\mathbf{F}_{4}$ is called a linear $[n,k]$ code over ${\bf
F}_{4}$ if $\mathcal{C}$ is a $k$-dimensional subspace of ${\bf
F}_{4}^{n}$. Note that for a linear $[n,k]$ code $\mathcal{C}$ over
$\mathbf{F}_{4}$ the number of codewords is $M=4^{k}$. Furthermore, we
call the linear code $\mathcal{C}$ a linear $[n,k,d]$ code if the
minimum distance of $\mathcal{C}$ is at least $d$. Let $A_i$ be the
number of codewords in $\mathcal{C}$ of Hamming weight $i$. The
sequence of numbers $A_0, A_1, \ldots, A_n$ is called the {\sl
weight distribution} of $\mathcal{C}$. It is well known in coding
theory\cite{ms} that for a linear code $\mathcal{C}$, the
distance distribution of $\mathcal{C}$ is equal to the weight
distribution of $\mathcal{C}$ and the minimum distance of $\mathcal{C}$ is
equal to the minimum Hamming weight of nonzero codewords. For fixed
$n$ and $k$, let $d_{4}(n,k)$ be the maximal minimum distance of a
linear $[n,k]$ code over $\mathbf{F}_{4}$. Tables of lower and upper
bounds on $d_4(n,k)$ are available in Ref.~\refcite{brou}.

We denote by
\begin{eqnarray}
\mathcal{A} & = & \{ \mathbf{x}=(x_1,x_2, \ldots, x_n) \in 
\mathbf{F}_{4}^{n}: x_i \not= 0 \; \mbox{for all} \; i \}  \nonumber \\
& = & \{ \mathbf{x}=(x_1,x_2, \ldots, x_n) \in \mathbf{F}_{4}^{n}: 
w(\mathbf{x})= n  \}
\end{eqnarray}
the set of vectors with maximal Hamming weight, that is: the set of words that
do not have the letter $A$.
For a linear $[n,k]$ code $\mathcal{C}$ over $\mathbf{F}_{4}$, by
(\ref{b2}), it is easy to check that
\begin{equation}
\bigcup_{\mathbf{c}\in \mathcal{C}} L(\mathbf{c}) 
= \mathcal{A} + \mathcal{C}\,.  
\end{equation}
Hence, by (\ref{b5}), the average error probability $\bar{e}$ can be
rewritten as
\begin{equation}
\bar{e} = 1- \frac{1}{3^{n}4^{k}} |\mathcal{A} + \mathcal{C}|\,. \label{d1}
\end{equation}
Note that $\mathcal{A} + \mathcal{C}$ is a union of some cosets of 
$\mathcal{C}$,
\begin{equation}
 \mathcal{A} + \mathcal{C} = \bigcup_{j=1}^{\alpha} (\mathbf{a}_j +
\mathcal{C})\,,\quad \mathbf{a}_j \in \mathcal{A}\,,
\label{coset}
\end{equation}
where $ \mathbf{a}_1 + \mathcal{C}, \mathbf{a}_2 + \mathcal{C}, \ldots, {\bf
a}_{\alpha} + \mathcal{C}$ are some different cosets of $\mathcal{C}$.
This implies that $|\mathcal{A} + \mathcal{C}|={\alpha}4^{k}$. Therefore,
the average error probability $\bar{e}$ is also given by
\begin{equation}
\bar{e} = 1- \frac{\alpha}{3^{n}}\,. \label{d2}
\end{equation}

\begin{Rmk}
For fixed $n$ and $k$, the best we can do is to choose
a linear $[n,k]$ code $\mathcal{C}$ over $\mathbf{F}_{4}$ such that
$\alpha$ is as large as possible. Note that even if $\mathcal{C}$ is
optimal in this sense, the average error probability may not be
small.
For fixed $n$ and $\bar{e}\leq \varepsilon$, we have to try to find
a linear $[n,k]$ code $\mathcal{C}$ over $\mathbf{F}_{4}$ with the
largest dimension $k$ such that
\begin{equation}
\alpha \geq 3^{n}(1-\varepsilon)\,.  
\end{equation}
Note that for certain values of $n$ and $\varepsilon$, 
such a linear code $\mathcal{C}$ may not exist.
\end{Rmk}

\begin{Rmk}
It is known from Ref.~\refcite{ck} (see Problem 11 on p.~114) 
that for our specific channel the codes in Shannon's Channel Coding Theorem 
can be replaced by linear codes over $\mathbf{F}_{4}$, that is, 
for any $0< \varepsilon <1$ and
\begin{equation}
\frac{k}{n}< \frac{1}{2}\log_{2}(4/3)\doteq 0.2075
\end{equation}
there will exist, for sufficiently large $n$, a linear $[n,k]$ code
$\mathcal{C}$ over $\mathbf{F}_{4}$ such that
\begin{equation}
\alpha \geq 3^{n}(1-\varepsilon)\,.  
\end{equation}
This is equivalent to the fact that $\bar{e}\leq \varepsilon$.
\end{Rmk}

In general, for a linear $[n,k,d]$ code $\mathcal{C}$ over $\mathbf{F}_{4}$,
by Theorem \ref{theo2} we have
\begin{equation}
\bar{e} \leq \frac{4^{k}-1}{2} \left(\frac{2}{3}\right)^d <
\frac{4^{k}}{2} \left(\frac{2}{3}\right)^d\,,
\label{d4}
\end{equation}
and if the decoding regions $D_1, D_2, \ldots, D_{4^{k}}$ are
given by (\ref{b3}), then by Theorem \ref{theo3},
\begin{equation}
e_{\max} \leq (4^{k}-1) \left(\frac{2}{3} \right)^d
< 4^k \left(\frac{2}{3} \right)^d\,.
\label{d5}
\end{equation}
Furthermore, if the weight distribution $\{A_s\}_{s=0}^{n}$ of
$\mathcal{C}$ is known, then by Theorem \ref{theo1},
\begin{equation}
\bar{e} \leq \frac{1}{2} \sum_{s=1}^n A_s
\left(\frac{2}{3}\right)^s. \label{d6}
\end{equation}

By using the Gilbert-Varshamov quasi-random construction of linear
codes,\cite{ms} one can construct, for sufficiently large $n$,
a linear $[n,k]$ code $\mathcal{C}$ over $\mathbf{F}_{4}$ with size
\begin{equation}
  M=4^{k} \doteq 4^{n[1-H_4(d/n)]} 
\end{equation}
such that the minimum distance $d(\mathcal{C}) \geq d$, where
\begin{equation}
  H_4(x) = x\log_{4}3 -x\log_{4}x -(1-x)\log_{4}(1-x)\,,\quad 0\leq x \leq
\frac{3}{4}\,.
\end{equation}
It follows from (\ref{d5}) that the maximum error probability $e_{\max}$ is
upper bounded by
\begin{equation}
e_{\max} < 4^k \left(\frac{2}{3} \right)^d
\doteq 4^{n[1-H_4(d/n)+(d/n)\log_{4}(2/3)]}\,.
\label{d3}
\end{equation}
The function of $d/n$ in the exponent is such that  
\begin{equation}
  1-H_4(x)+x\log_{4} \frac{2}{3} <0 \Longleftrightarrow x> \beta\,,
\end{equation}
where $\beta \doteq 0.4627$ is the unique solution of
$1-H_4(x)+x\log_4 (2/3)=0$.
This means that, for sufficiently large $n$,
one can construct a linear $[n,k]$ code $\mathcal{C}$ over $\mathbf{F}_{4}$ 
with the rate
\begin{equation}
  \frac{k}{n} \doteq 1-H_4(\beta) \doteq 0.1353 
\end{equation}
such that the maximum error probability $e_{\max}$ is arbitrarily small.

\section{Some Examples} \label{sec8}

In this section we give some examples of linear codes for the specific channel
in Section \ref{sec3} to illustrate our results. These linear codes
are listed in Brouwer's tables\cite{brou} of presently best known
quaternary linear codes. The value of $R$ is obtained from
$M=2^{nR}$ in Theorem \ref{theo-shannon} and corresponds to the 
efficiency mentioned in Section \ref{sec1}.

\begin{table}[h]
\tbl{Examples of linear codes for the channel specified by the conditional
  probability distribution (\ref{a7}).
  For each code we give the number $n$ of letters in each codeword, the
  dimension $k$ of the subspace of $\mathbf{F}_4^n$, the minimum Hamming
  distance $d$, the total number $M$ of codewords, the efficiency $R$, and in
  the last column an upper bound of (\ref{d4}) on the average error
  probability $\bar{e}$, rounded to four significant digits. 
  \newline 
  The first group on the left are six  codes of length $n=100$ with
  consecutive values of the dimension~$k$.
  For $k\geq 16$, the upper bound on $\bar{e}$ is greater than $1$, 
  and so it is not meaningful.
  The second group on the left are two codes of large lengths which
  demonstrate that the error probability can be made very small.
  The group on the right are 14 codes with lengths decreasing from $50$
  to $10$.
\label{tbl}}
{\begin{tabular}{@{}rrrlllcrrrrll@{}} \toprule
\multicolumn{1}{c}{$n$} &
\multicolumn{1}{c}{$k$} &
\multicolumn{1}{c}{$d$} &
\multicolumn{1}{c}{$M$} &
\multicolumn{1}{c}{$R$} &
\multicolumn{1}{c}{$\bar{e}\leq$} &\quad&
\multicolumn{1}{c}{$n$} &
\multicolumn{1}{c}{$k$} &
\multicolumn{1}{c}{$d$} &
\multicolumn{1}{c}{$M$} &
\multicolumn{1}{c}{$R$} &
\multicolumn{1}{c}{$\bar{e}\leq$} \\
\colrule
100 & 10 & 62 & $4^{10}$ & 0.2 & $6.337\times 10^{-6}$ &&
 50 &  5 & 35 & $1024$   & 0.2 & $3.516\times10^{-4}$\\
100 & 11 & 60 & $4^{11}$ & 0.22 & $5.704\times 10^{-5}$ &&
 50 &  6 & 33 & $4096$   & 0.24 & $3.165\times10^{-3}$\\
100 & 12 & 58 & $4^{12}$ & 0.24 & $5.133\times 10^{-4}$ &&
 48 &  6 & 32 & $4096$   & 0.25 & $4.747\times10^{-3}$\\
100 & 13 & 56 & $4^{13}$ & 0.26 & $4.620\times 10^{-3}$ &&
 48 &  5 & 33 & $1024$   & 0.208 & $7.912\times10^{-4}$\\
100 & 14 & 55 & $4^{14}$ & 0.28 & $0.02772$ &&
 47 &  6 & 31 & $4096$   & 0.255 & $7.120\times10^{-3}$\\
100 & 15 & 52 & $4^{15}$ & 0.30 & $0.3742$ &&
 46 &  5 & 32 & $1024$   & 0.217 & $1.187\times10^{-3}$\\
&&&&&&&
 45 &  5 & 31 & $1024$   & 0.222 & $1.780\times10^{-3}$\\
&&&&&&&
 43 &  5 & 30 & $1024$   & 0.233 & $2.670\times10^{-3}$\\
&&&&&&&
 42 &  5 & 29 & $1024$   & 0.238 & $4.005\times10^{-3}$\\
200 & 20 &109 & $4^{20}$ & 0.2   & $3.517\times10^{-8}$ &&
 41 &  5 & 28 & $1024$   & 0.244 & $6.008\times10^{-3}$\\
250 & 25 &136 & $4^{25}$ & 0.2   & $6.340\times10^{-10}$ &&
 40 &  4 & 28 & $256$    & 0.2 & $1.502\times10^{-3}$\\
&&&&&&&
 30 &  3 & 22 & $64$   & 0.2 & $4.277\times10^{-3}$\\
&&&&&&&
 20 &  2 & 16 & $16$   & 0.2 & $0.01218$\\
&&&&&&&
 10 &  1 & 10 & $4$   & 0.2 & $0.02601$\\
\botrule
\end{tabular}}
\end{table}

In Table~\ref{tbl} we give the parameters of various codes from 
Ref.~\refcite{brou}.
They illustrate a general observation, namely that there is a trade-off
between the simplicity of the code (short length $n$ of the words, 
small number $M$ of them) on one side and the performance of the code (large
efficiency $R$, small average error $\bar{e}$). 
In addition to the codes of Table~\ref{tbl}, we mention the following four
codes of moderate length and reasonably good performance. 

\begin{Epl} Here we list explicitly known linear codes of
moderate lengths.
\begin{tabular}{@{}p{18pt}@{}p{342pt}@{}}
(1) & A code with $n=28\,,\ k=4\,,\ d=20\,, M=256$ for which 
      $R=\frac{2}{7}\doteq0.2857$ and 
      $\bar{e}< \frac{4^{4}}{2}(2/3)^{20}\doteq 0.03849$ 
      is the upper bound of (\ref{d4}).\newline
As is known from Ref.~\refcite{brou}, there exists a linear $[28,4,20]$
code over $\mathbf{F}_{4}$ with the weight distribution
$ A_{20}=189\,,\ A_{24}=63\,,\ A_{28}=3$, so that (\ref{d6}) gives the bound 
$\bar{e}\leq 0.03038$.
\end{tabular}
\begin{tabular}{@{}p{18pt}@{}p{342pt}@{}}
(2)& A code with $n=31\,,\ k=4\,,\ d=22\,,\ M=256$ for which 
     $R=\frac{8}{31}\doteq0.2581$ and 
     $\bar{e}<\frac{4^{4}}{2}(2/3)^{22}\doteq 0.01711$ is
      the upper bound of (\ref{d4}). 
     \newline
As is known from Ref.~\refcite{brou}, there exists a linear $[31,4,22]$
code over $\mathbf{F}_{4}$ with the weight distribution
$ A_{22}=141\,,\ A_{24}=87\,,\ A_{28}=24\,,\ A_{30}=3$, 
so that (\ref{d6}) gives the bound $\bar{e}\leq 0.01216$.
\end{tabular}
\begin{tabular}{@{}p{18pt}@{}p{342pt}@{}}
(3)& A code with  $n=40\,,\ k=5\,,\ d=28\,,\  M=1024$ for which
     $R=\frac{1}{4}=0.25$ and 
     $\bar{e}<\frac{4^{5}}{2}(2/3)^{28}\doteq 0.006008$
    is the upper bound of (\ref{d4}). \newline
As is known from Ref.~\refcite{brou}, this optimal linear code is a
quasi-cyclic code. The generator matrix can be represented as
$G=[G_0, G_1, G_2, G_3, G_4, G_5, G_6, G_7]$
where $G_i$ for $0\leq i \leq 7$ are $5\times 5$ circulant matrices. 
The first row of $G$ is given by
$[10000 \quad 10120 \quad 11020 \quad 11230 \quad 12220 \quad 13130
\quad 13210 \quad 11312 ]\,,$
where we identify $2$ with $a\in\mathbf{F}_4$
and $3$ with $b\in\mathbf{F}_4$.
\end{tabular}
\begin{tabular}{@{}p{18pt}@{}p{342pt}@{}}
(4) & The shortened code of the previous example:
$n=39\,,\ k=4\,,\ d=28\,,\ M=256$ for which $R=\frac{8}{39}\doteq0.2051$ 
and $\bar{e}<\frac{4^{4}}{2}(2/3)^{28} \doteq 0.001502$.
\end{tabular}
\end{Epl}

\begin{Rmk}
For codes of small size, one can calculate the exact values
of $e_{i}$, $\bar{e}$,
and $e_{\max}$ by using (\ref{b4})--(\ref{b7}). The decoding method
is also computationally feasible.
For codes of large size, for example $M=4^{20}$,
decoding will become an enormous computational task.
\end{Rmk}

\begin{Rmk}
By using nonlinear codes, it may be possible to achieve better results
on the decoding error probability,
but we have not tried to search for good quaternary nonlinear codes in
the literature.
\end{Rmk}

\section*{Acknowledgments}
This research is supported in part by the DSTA research grant
R-394-000-011-422 and in part by ICITI research grant R-144-000-109-112. 
The work of Fang-Wei Fu is also supported by the
National Natural Science Foundation of China (Grant No.\ 60172060),
the Trans-Century Training Program Foundation for the Talents by the
Education Ministry of China, and the Foundation for University Key
Teacher by the Education Ministry of China.

\end{document}